# Low temperature ferromagnetic properties, magnetic field induced spin order and random spin freezing effect in $Ni_{1.5}Fe_{1.5}O_4$ ferrite; prepared at different pH values and annealing temperatures

R.N. Bhowmik[*] and K.S. Aneeshkumar

Department of Physics, Pondicherry University, R. Venkataraman Nagar,

Kalapet, Pondicherry-605014, India

[*]Corresponding author: Tel.: +91-9944064547; Fax: +91-413-2655734

E-mail: rnbhowmik.phy@pondiuni.edu.in

**ABSTRACT**

We present the low temperature magnetic properties in $Ni_{1.5}Fe_{1.5}O_4$ ferrite as the function of pH at which the material was prepared by chemical route and post annealing temperature. The material is a ferri/ferromagnet, but showed magnetic blocking and random spin freezing process on lowering the measurement temperature down to 5 K. The sample prepared at pH ~12 and annealed at 800 $^0$C showed a sharp magnetization peak at 105 K; the superparamagnetic blocking temperature of the particles. The magnetization peak remained incomplete within measurement temperature up to 350 K for rest of the samples, although peak temperature was brought down by increasing applied dc field. The fitting of temperature dependence of coercivity data according to Kneller′s law suggested random orientation of ferromagnetic particles. The fitting of saturation magnetization according to Bloch′s law provided the exponent that largely deviated from 3/2, a typical value for long ranged ferromagnet. An abrupt increase of saturation magnetization below 50 K suggested the active role of frozen surface spins in low temperature magnetic properties. AC susceptibility data elucidated the low temperature spin freezing dynamics and exhibited the characters of cluster spin glass in the samples depending on pH value and annealing temperature.

Key words: Ferrite nanoparticles, random spin freezing, ac susceptibility, cluster spin glass.

## 1. INTRODUCTION

Spinel ferrites, having general formula of $MFe_2O_4$ (M = divalent metal ion, e.g. Ni, Co, Cu, etc.), remained as one of the most attractive magnetic oxides due to a rich class of magnetic and electronic properties. The interest for Nickel ferrite ($NiFe_2O_4$) stems from the fact that it is a soft ferri/ferromagnet with reasonably high magnetization. The magnetic properties of $NiFe_2O_4$ are determined primarily by the superexchange interactions between Ni and Fe ions in tetrahedral (A) and octahedral (B) sites of cubic spinel lattice structure [1-2]. Without elemental substitution effect, the long-ranged ferrimagnetic spin order in $NiFe_2O_4$ can also be perturbed in nanocrystalline form by surface effects (spin disorder, anisotropy, and exchange interactions) and exchange of Ni and Fe ions among the A and B sites. This increases multi-functionalities in nanocrystalline form of soft ferromagnetic ferrites and make them suitable for applications in the field of power electronics (transformer core), recording media, microwave devices, and medical treatment (hyperthermia, ferrofluid) [3-5]. The tuning of ferromagnetic parameters (coercivity, saturation magnetization, magnetic blocking, and ferromagnetic squareness) in a wide temperature scale is important for applying the magnetic material. Hence, it is necessary to understand the basic mechanisms that control the tuning of ferromagnetic parameters in magnetic materials.

From literature reports, it is understood that the nature of surface spin structure, magnetic frustration and inter-particle interactions determine the low temperature magnetic properties of nanoparticles in the form of superparamagnetic blocking, spin glass, and exchange bias effect [6-9]. The reduction of magnetization, higher coercivity, and surface spin disorder in nanoparticles of $NiFe_2O_4$ are remarkably different from that in micron-sized particles [2, 10]. Core-shell spin structure has been modeled to explain the magnetic reduction in $NiFe_2O_4$ nanoparticles [1, 11], where spins in the core are ferrimagnetically aligned as in bulk and shell forms a disordered spin structure that controls the magnetic reduction and exhibition of spin glass and superparamagnetic feature. The exchange of Ni and Fe ions among A and B sites and surface spin canting were identified as other factors that determine magnetic reduction and spin glass feature in $NiFe_2O_4$ nanoparticles [1, 9]. Lee et al. [12] considered the intermediate spin layer between core and shell for the magnetic field induced shift of blocking temperature. It is a great challenge to distinguish the blocking of core spins from the freezing of surface spins [6]. Kodama et al. [11] attributed the spin glass like feature in $NiFe_2O_4$ nanoparticles due to freezing of frustrated surface spins. In NiO nanoparticles, Winkler et al. [13] proposed that antiferromagnetic core of NiO was blocked at higher temperature and disordered surface spins showed spin glass like freezing at low temperature. Recent studies have shown that blocking/freezing phenomena and ferromagnetic properties in nanoparticles remarkably depend on the variation of synthesis conditions (pH value, annealing temperature, annealing atmosphere, etc) of chemical routed ferrite samples [9-10, 14].

We reported the room temperature magnetic properties of $Ni_{1.5}Fe_{1.5}O_4$ ferrite [15] with the variation pH value and annealing temperature. The as-prepared material was annealed in the temperature range 500-1000 $^0C$ [16]. Using dielectric spectroscopy study [17], we observed a transformation in electrical charge dynamics from low temperature semiconductor state to high temperature semiconductor state with an intermediate metal like state in the samples. The metal like state was understood as an effect of the crossover of localized hopping of electronic charge (electrons) at low measurement temperatures to thermally activated long range hopping at higher temperatures. In this work, we report the electronic spin dynamics in $Ni_{1.5}Fe_{1.5}O_4$ samples under magnetic field for measurement temperature down to 5 K. Our objective is to study the effects of the variation of pH value during chemical reaction and post annealing temperature on low temperature magnetic phenomena (blocking/freezing of electronic spin moment) and tuning of ferromagnetic parameters in $Ni_{1.5}Fe_{1.5}O_4$ ferrite.

## 2. EXPERIMENTAL

### 2.1 Sample preparation

Details of the material preparation and characterization were reported earlier [15-17]. The samples were prepared by chemical reaction of the stoichiometric amount of $Ni(NO_3)_2.6H_2O$ and $Fe(NO_3)_3.9H_2O$ salts in solution at 80 °C by maintaining pH at 6, 8, and 12. Finally, the chemical routed (as-prepared) material was made into pellets and annealed at selected temperatures. The X-ray diffraction pattern ($CuK_\alpha$ lines with λ =1.5406 A°) and Raman spectra were used to confirm the formation of single phased cubic spinel structure. The material chemically reacted at pH value 8-12 formed single phase upon annealing the as-prepared material in air. The sample prepared at pH 6 formed a minor amount of α-$Fe_2O_3$ when annealed in air. However, the sample when annealed under vacuum (~$10^{-5}$ mbar) formed single-phased cubic spinel structure with space group Fd3m. The samples are labeled as NFpHX_Y, where X is pH and Y is annealing temperature in degree centigrade. The structural information (lattice parameter and grain size) of the single-phased samples used in this work is indicated in Fig. 1 for information to readers.

### 2.2 Measurement

Magnetic properties of the samples were studied using physical properties measurement system (PPMS-EC2, Quantum Design, USA). The temperature dependent dc magnetization (M) was measured using zero field cooling (ZFC) and field cooling (FC) modes. In ZFC mode, the sample was cooled from higher temperature (320 K) without applying external magnetic field down to the lowest measurement temperature (5 K). Then, magnetic measurement started in the presence of set magnetic field (say, 100 Oe) and MZFC(T) data were recorded during the increase of temperature (T) from 5 K to 300 K/320 K. In FC mode, the sample was cooled under set magnetic field from 320 K to low temperature (5 K) and MFC(T) data were recorded without changing the magnetic field during the increase of temperature up to 320 K. The magnetic field (H) dependence of dc magnetization (M(H)) was measured by zero field cooling the sample from 320 K to the measurement temperature, which was set in the range 5-300 K. The ac susceptibility (real: $\chi'$ and imaginary: $\chi''$ components) data were recorded in the temperature range 10 K-340 K by applying ac magnetic field (amplitude 1 Oe with frequencies in the range 37 Hz -10 kHz).

## 3. Results and discussion

### 3.1. *Temperature dependent magnetization*

Fig. 1 shows the temperature dependence of MZFC and MFC curves at dc field of 100 Oe for the samples, chemically synthesized at pH values 6 (Fig. 1(a-d)), 8 (Fig.1 (e-g)), 12 (Fig.1 (h-i)) and annealed at different temperatures. The samples, prepared at pH 6 and 8, exhibited the characters of ferro/ferrimagnetic nanoparticles with splitting between MFC(T) and MZFC(T) curves below 320 K, where MZFC(T) curve decreased and MFC(T) curve increased on decreasing the measurement temperature down to 5 K. The MZFC (T) curves of these samples indicated a broad maximum or incomplete maximum within the measurement temperature limit 320 K. However, magnetic gap between MZFC and MFC curves at lower temperature decreases and the peak appears to be shifted to higher temperature on increasing the annealing temperature (and increase of grain size) of the samples. The magnetization of the samples prepared at pH 6 is found higher than the samples prepared at pH 8 and 12. An additional magnetic shoulder is clearly appeared at low temperature (below 30 K) for the samples prepared at pH 6 and annealed at higher temperature (800-1000 °C). The origin of low temperature magnetic feature will be discussed using the ac susceptibility data. A different type of magnetic feature is observed for the samples prepared at pH 12 and annealed at 800 $^0$C (grain size ~ 6 nm). This sample exhibited a well defined superparamagnetic blocking temperature ($T_B$) at about 105 K, and splitting between MZFC and MFC curves below the blocking temperature (Fig. 1(h)). On increasing the annealing temperature to 1000 $^0$C, the sample prepared at pH 12 was not able to achieve the blocking temperature within 300 K (Fig. 1(i)). This is the effect of increasing grain size (6 nm to 29 nm) in the material. From physics point of view, the FC curve is in quasi-equilibrium state due to local ordering of spins or cluster of spins during field cooling process while the ZFC curve is in non-equilibrium blocking state when relaxation time ($\tau$) of the spins or cluster of spins is greater than the magnetic measurement time ($\tau_m \sim 10^2$ s). Since MZFC(T) curves exhibited a broad maximum or incomplete maximum, it is difficult to exactly determine the blocking temperature for most of the samples. A broad peak indicates a distribution of blocking temperature ($T_B$), which can be related to distribution of magnetic anisotropy constant (K) and grain volume (V) by the relation $k_B T_B \sim 25KV$ [18]. In such case, the temperature derivative of MZFC(T) curve ($\frac{dMZFC}{dT}$) was fitted with Gaussian shape to get the information of the distribution of relaxation time or anisotropy barrier of the magnetic particles below $T_B$ [19]. The average blocking temperature was estimated from the intercept of $\frac{dM_{ZFC}}{dT}$ vs. T curves on temperature axis where $\frac{dMZFC}{dT} = 0$. The fit parameters, like peak position (inflection point of $M_{ZFC}$ (T) curves below $T_B$), full width at half maximum (FWHM) of the peak, and peak height) were used as the initial input parameters to define a distribution curve with median blocking temperature ($T_{Bm}$) and $T_{Bm}$ is found near to inflection point ($T_P$) of the MZFC(T) curve below ($T_B$). The distribution curve was fitted with log-normal distribution ($f(t_{Bm})$) function.

$$f(t_{Bm}) = \frac{1}{\sqrt{2\pi}\sigma} \frac{1}{T_{Bm}} exp\left(-\frac{ln^2 t_{Bm}}{2\sigma^2}\right) \qquad (1)$$

Accuracy of the obtained parameters were checked by a quantitative analysis of the MZFC(T) curves using the following equation [18, 20], where $f(t_{Bm})$ has been replaced by $f(t_B)$.

$$M_{ZFC}(T) = C + \frac{M_{Sat}^2 H}{3K_{eff}}\left[ln\left(\frac{\tau_m}{\tau_o}\right)\int_0^t \frac{t_B}{t}f(t_B)dt_B + \int_0^t f(t_B)dt_B\right] \qquad (2)$$

C is a constant that takes into account the residual value of low temperature magnetization. The first term inside the bracket corresponds to the contribution from mutually interacting super-paramagnetic nanoparticles with the factor $ln\left(\frac{\tau_m}{\tau_o}\right)$ is 25. The second term corresponds to the contribution from superparamagnetic particles in relaxed state ($\tau > \tau_m$). $M_{sat}$ is the saturation magnetization (the value is taken from M(H) curves at 5 K/10 K); H is the applied magnetic field during ZFC measurement; $K_{eff}$ is the effective anisotropy constant, $t$ is the reduced temperature ($T/T_{Bm}$) and $t_B$ is the reduced blocking temperature ($T_B/T_{Bm}$). The example of typical distribution (f($t_{Bm}$)) curve and final fit of the respective MZFC(T) curve are shown for samples NFpH8_500 (Fig. 1(e)) and NFpH12_800 (Fig. 1(h)).

The total anisotropy energy per unit volume (E) of a magnetic particle in the presence of external magnetic field (H) and at specific measurement temperature (T) below the blocking temperature (T < $T_B$) is governed by the competition between uniaxial anisotropy energy ($E_A = K_{eff}\sin^2\theta$) and Zeeman energy ($E_H = -M_{sat}H\cos(\theta - \varphi)$) [21], as defined below.

$$E(T, H) = K_{eff}(T)\sin^2\theta - M_{sat}(T)H\cos(\theta - \varphi) \qquad (3)$$

θ is the angle between magnetization vector and local anisotropy easy axis (EA), and φ is the angle between H and EA. At higher fields, the increase of Zeeman energy will reduce the effective anisotropy barrier for blocking of the particle and subsequently, $T_B$ will shift to lower values. Fig. 2(a-c) demonstrates the applied field induced shift of $T_B$ (reduction) for the samples NFpH6_600 (Fig.2(a), NFpH8_500 (Fig.2(b)), and NFpH12_800 (Fig.2(c). First order derivative of the MZFC(T, H) curves and final distribution curves used for fitting of the field dependence of MZFC(T) data are shown for NFpH8_500 (Fig. 2(d-e)) and NFpH12_800 (Fig. 2(f-g)) samples. Fig. 3 shows the variation of peak parameters (position: Tp, peak height, and FWHM) from Gaussian fit of dMZFC(T)/dT curves with the applied field increment. The general features are as follows, (1) the peak position of dMZFC(T)/dT curves shifts to lower temperature with the increase of applied magnetic field and it can be taken as equivalent information of the field dependence of $T_B$, (2) the peak height increases with the increase of applied field, and (3) the peak width decreases with the increment of applied field. The increase of peak height with the decreasing width confirms the magnetic field induced clustering of small magnetic domains/ particles to form a larger sized multi-domain particle. This leads to narrowing of the distribution curve with reduced height, indicating the magnetic field induced reduction of anisotropy barrier and magnetic exchange interactions in the system [19]. The fact is supported by the reduction of area (height and width) of distribution curves (Fig. 2(e, g)) with the increase of applied magnetic field. In fitting of MZFC(T) data using equation (2), we have used the low temperature $M_{sat}$ of the samples (34.1 emu/g, 30.1 emu/g, 16.5 emu/g for NFpH6_600, NFpH8_500, NFpH12_800, respectively). In Table 1, we have shown the constant *C* and the anisotropy constant per unit field ($K_{eff}/H$) values required to fit the MZFC(T) curves at different magnetic field. We observed that fit of the MZFC(T) curves are not much accurate for NFpH6_600 sample, exhibited more peak broadness. However, $K_{eff}/H$) values for NFpH8_800 and NFpH12_800 samples decreased within error bar (not shown) with the increase of field magnitude. This takes into account the effect of decreasing distribution curve area on increasing magnetization at higher field. Secondly, width and $T_p$ for the NFpH8_500 sample (grain size ~ 10 nm) are higher in comparison to the values in NFpH12_800 sample (grain size ~ 6 nm). The results show that magnetic distribution in superparamagnetic blocking regime is affected by grain size variation; exhibiting a narrow distribution of anisotropy barriers for the samples prepared at low annealing temperature and a large distribution for the samples annealed at higher temperature. Fig. 3(d) illustrates the magnetic field induced reduction of $T_B$, estimated from $\frac{dM_{ZFC}}{dT} = 0$ on temperature axis, for the samples NFpH6_600, NFpH8_500, and NFpH12_800. In order to identify the nature of magnetic order below $T_B$, whether spin glass or superparamagnetic blocking or random spin freezing of magnetic particles, the variation of $T_B$ with applied field has been fitted with power law [22]: $T_B(H) = a - bH^n$ with field exponent ($n$) ~ 0.18, 0.14, and 0.22 for NFpH6_600, NFpH8_500, and NFpH12_800 samples, respectively (Fig. 3(e)). The fitting of field dependence of mean blocking temperature ($T_{Bm}(H)$) gives the exponent ($n$) ~ 0.12 and 0.257 for NFpH8_500 and NFpH12_800, respectively. As discussed in Ref. [22-24] and references therein, the $n$ is expected ~ 0.67 according to Almeida-Thouless line for classical spin glass or typical superparamagnetic nanoparticles with single relaxation time (Neel-Brown model). The typical $n$ value for system of cluster spin glass coexists with ferromagnetic order (e.g., $La_{0.5}Sr_{0.5}CoO_3$) is ~ 0.58. The $n$ value is ~ 0.48 for 3D Ising spin glass with random anisotropy. The exponent ($n$) values in our samples ($n$ = 0.14-0.22) are much smaller than the above specific class of disordered magnetic systems. The field exponent value covered a wide range 0.023-0.46 for the $Co_{0.2}Zn_{0.8}Fe_{1.95}Ho_{0.05}O_4$ ferrite, where spin glass and superparamagnetism are coexisting with ferrimagnetic order below the magnetic blocking temperature. Hence, the divergence between MZFC and MFC curves below $T_B$ in $Ni_{1.5}Fe_{1.5}O_4$ ferrite samples can be attributed to a random freezing of spins or relaxation of magnetic clusters with a strong intra-cluster interaction along local anisotropy axes. The field induced magnetic spin ordering below room temperature is now demonstrated from the magnetic field dependence of magnetization (M(H)) measurements.

*3.2 Field dependent magnetization*

M(H) data were measured under ZFC mode at selected temperatures in the temperature range 5 K-300 K by sweeping the magnetic field within ± 60 kOe. However, M(H) data are shown within magnetic field range ± 10kOe for clarity of the information. Fig. 4(a-d) shows a comparative plot of M(H) data measured at 10 K and 300 K for the samples prepared at pH 6 and thermally annealed at different temperatures. The immediate observation is that the samples at 10 K showed a wide loop in comparison to a narrow loop at 300 K, indicating a transformation of features from medium hard magnet at low temperature to soft ferromagnetic character at room temperature. Additionally, the soft ferromagnetic character is rapidly enhanced (with increasing magnetization) on increasing annealing temperature of the material chemically prepared at pH 6. Such samples are useful for applications in transformer core and hyperthermia [21]. We have not seen any appreciable shift of the M(H) loop measured under field cooling @70 kOe with respect to the M(H) loop measured under ZFC mode. A typical example is shown for NFpH6_600 at 10 K, which, of course, showed a minor enhancement of positive magnetization after field cooling. The M(H) data at selected temperatures are shown for NFpH6_600 (Fig. 4(e)), NFpH8_500 (Fig. 4(f)), and NFpH12_800 (Fig. 4(g)) samples. As shown in Table 1, the ferromagnetic parameters (magnetization, squareness, coercivity, anisotropy constant) of the samples reduced with the increase of measurement temperature. The samples NFpH6_600 and NFpH8_500 clearly showed a loop at all measurement temperatures, where as NFpH12_800 does not show magnetic loop and retaining ($M_R$) of high field magnetization after reducing the field to zero for temperatures at $\geq$ 200 K. The magnetic squareness (S), defined as the ratio of $M_R$ and spontaneous magnetization ($M_S$), is an important parameter to estimate the retaining of high field magnetization. Magnetic squareness of the present system is reasonably good and it increased up to the value 0.28-0.46 on lowering the temperature down to 5 K. In case of uniaxial anisotropy in magnetic material, the squareness is expected close to 0.5 [25]. The small smaller value of squareness in NFpH12_800 can be attributed to surface spin disorder effects. In the absence of magnetic saturation at higher fields (up to 60 kOe), the $M_s$ (spontaneous magnetization) was calculated using Arrot plot ($M^2$ vs. H/M) [15] of initial M(H) curve and it is represented in the insets of Fig. 4(f-g). The lack of magnetic saturation indicates the randomly distributed magnetic spins structure and exchange interactions in nanoparticles. The inset of Fig. 4(h) shows that spontaneous magnetization in the material at 5 K has increased with annealing temperature. It is worthy to mention that $M_S$ at low temperatures in our Ni rich ferrite samples is comparable to the reported value for nickel ferrite nanoparticles [1, 25]. For NFpH12_800 sample, $M_s$ at 5 K (13.2 emu/g) is relatively low due to coexistence of a significant fraction of superparamagnetic component along with ferromagnetic component in smaller grain sized sample [26]. It may be noted (Table 1) that $M_s$ of the samples at different measurement temperatures are smaller than the high field magnetization. The saturated magnetization ($M_{sat}$) has been determined from the fitting of high-field M(H) curves using the law of approach to saturation of magnetization [19].

$$M(H) = M_{sat}\left(1 - \frac{a}{H} - \frac{b}{H^2}\right) + \chi_d H \qquad (4)$$

Here, $\chi_d$ is the high field induced paramagnetic susceptibility, $a$ and $b$ are the constants. The term $\frac{a}{H}$, which takes into account the existence of micro-structural defects in the system, is generally negligible [27]. The $K_{eff}$ was calculated using the relation $b = \frac{8}{105} \times \left(\frac{K_{eff}}{M_{sat}}\right)^2$ [14]. M(H) curves of the samples NFpH6_600, NFpH8_500 and NFpH12_800 were fitted with equation (4) and shown in Fig. 4(d) for NFpH6_600 sample. Table 1 shows the ferromagnetic parameters obtained from fitting of M(H) curves using equation (4). Among the three samples, $K_{eff}$ values are relatively high in NFpH8_500 and low in NFpH12_800. Although equation (4) is applicable for NFpH12_800 sample, where superparamagnetic component plays a major role on determining the magnetic properties, its M(H) curves are best fitted with the replacement of $\chi_d H$ in equation (4) by Langevin function for superparamagnetic component of magnetization [28].

$$M(H) = M_{sat\_f}\left(1 - \frac{b}{H^2}\right) + M_{sat\_sp}\left(\coth\left(\frac{\mu H}{k_B T}\right) - \left(\frac{k_B T}{\mu H}\right)\right) \qquad (5)$$

The $M_{sat\_f}$ and $M_{sat\_sp}$ are contributions from ferromagnetic and superparamagnetic components to the saturated magnetization. The contribution from ferromagnetic component ($M_{sat\_f}$ ~16.46 at 5 K and 6.15 emu/g at 300 K) to saturated magnetization of NFpH12_800 sample is nearly same ($M_{sat}$) as got from fit of M(H) curves using equation (4). The contribution of superparamagnetic component ($M_{sat\_sp}$), as plotted in the inset of Fig. 5(a), is remarkably high for NFpH12_800 sample. Interestingly, saturated magnetization rapidly increased below 50 K. It can be attributed to an additional high field magnetic contribution from randomly frozen surface spins at low temperatures [9, 27]. Otherwise, saturated magnetization decreases above 50 K. The main reason of the decreasing magnetization can be attributed to the breaking of ground state ferro/ ferrimagnetic spin order by low energy excitations of the spins in core as well as in disordered surface of ferromagnetic nanoparticles. This is realized by fitting the temperature dependence of saturated magnetization data using Bloch′s law [27, 29-30].

$$M_{sat}(T) = M_{sat}(0)(1 - BT^\alpha) \qquad (6)$$

$B$ is the Bloch constant. $M_{sat}(0)$ is the saturation magnetization at 0 K. α is the exponent that is dependent on magnetic spin order and typically 3/2 for long ranged ferromagnet. The constant $B$ was reported ~ $10^{-4}$-$10^{-5}$ for nanoferrites and ~ $10^{-6}$ for bulk ferromagnets. In bulk ferromagnetic ferrites, α is reported ~ 2 or less (e.g., ~ 2 for $CoFe_2O_4$, $NiFe_2O_4$, $Fe_3O_4$, and ~ 1.5 for $MnFe_2O_4$) [27, 31]. In nanomaterials, the temperature dependent magnetization deviated from Bloch′s law with exponent α = 1.5, which varies in a wide range 0.58-2.44 [9, 25, 30, 32]. The Bloch′s law is valid in our samples for measurement temperature above 50 K. In NFpH12_800 sample, the $M_{sat\_f}(T)$ data are fitted with α ~1.39 and $B$~1.94x$10^{-4}$. However, $M_{sat\_sp}(T)$ data do not follow Bloch′s law (inset of Fig. 5 (a)). On the other hand, $M_{sat}(T)$ data of NFpH6_600 and NFpH8_500 samples followed Bloch's law with noticeably large values of α ~ 2.73 and 2.89 with $B$ ~2.25x$10^{-8}$ and 8.65x$10^{-9}$, respectively. Our results suggest that magnetic spin interactions are sufficiently strong for the NFpH6_600 and NFpH8_500 samples with larger grain size and the spin interaction is highly perturbed for the NFpH12_800 sample with smaller grain size. The magnetic upturn at low temperature deviated from the Bloch law and it is termed as the quantization of spin-wave spectrum in nanoparticles, where surface spin freezing [30] or site exchange of cations [9] plays an important role. Table 1 showed a monotonic decrease of coercivity ($H_c$) with the increase of measurement temperature. Fig. 5(b) shows a good fit for $H_c(T)$ data with Kneller's law: $H_c(T) = H_c(0)[1 - (\frac{T}{\beta T_B})^k]$ [14, 28, 33]. The best linear fit of the $H_c(T)$ data showed k ~ 0.5 for NFpH6_600, k ~ 0.41 for NFpH8_500 and k ~ 0.21 for NFpH12_800, respectively. The obtained *k* values are lying within the theoretically predicted range of 0-1.5 for randomly oriented magnetic nanoparticles [21]. The shape of M(H) loop also suggest that magnetic particles in our samples are randomly oriented, where φ in equation (3) is non-zero [40]. We observed two important points regarding micro-structural change, viz., a decrease of grain size of the material with the increase of pH value during chemical reaction, and an increase of grain size during increase of annealing temperature of the chemical routed material. Such micro-structural changes are significantly affecting the magnetic spin dynamics of particles, especially at surface spins. The coercivity ($H_c$) of magnetic particles at low temperature (5K/10K) has a noticeable dependence on grain size (D). It followed (Fig. 5(b)) the relation of $H_C(D) = a + \left(\frac{b}{D}\right)$, proposed for multi-domain state of magnetic particles [10, 27]. The fitted values of constants (*a* and *b)* are tabulated in Fig. 5. The fit of $H_C(D)$ data at low temperature is characteristically different from the annealing temperature (grain size) dependence of coercivity measured at room temperature (300 K) [15]. One possibility is that small-sized grains (magnetic domains) of the samples are getting clustered at low measurement temperature and responds like from multi-domain particles. The $K_{eff}$ vs. 1/D plot (Fig. 5(c)) using low temperature data (5 K/10 K) followed an empirical relation $K_{eff} = K_v + \left(\frac{6}{D}\right)K_s$ [27, 34], where $K_v$ and $K_s$ represent anisotropy contributions from (core) bulk and surface parts of the nanoparticles. We have found negative intercept ($K_v$) with positive slope (6 $K_s$) for relatively large grain size with multi-domain nature (regime 1) and it becomes reverse (*$K_v$ is positive. slope is negative*) for smaller grain size with single domain nature (regime 2) of the samples. In regime 1, the effective anisotropy constant ($K_{eff}$), i.e., anisotropy energy per unit volume/gram of the material, increased due to increasing surface anisotropy contribution with the decrease of grain size. In regime 2, $K_{eff}$ is decreased from $K_v$ due to decrease of surface anisotropy contribution (increase of spin disorder) on decreasing the grain size in single domain range of magnetic particles (Fig. 5(d)). The magnitude of $K_v$ lies in the range (3.0-4.8)x$10^4$ emu Oe/g in regime 1 and in the range (2.3-12) x$10^5$ emu Oe/g for regime 2. Similarly, the magnitude of $6K_s$ lies in the range (1.7-19.8)x$10^6$ nm-emu Oe/g for regime 1 and in the range (1.4-11.4) x$10^6$ nm-emu Oe/g for regime 2.

*3.3 Magnetic dynamics using AC susceptibility analysis*

The spin freezing phenomena has been studied from the real $\chi'(T)$ and imaginary $\chi''(T)$ parts of ac susceptibility in the measurement temperature range 10 K-340 K for selected samples. The $\chi'(T)$ and $\chi''(T)$ data (Fig. 6(a-b)) for NFpH6_500 sample showed dispersion on increasing the driving frequency (*f*) from 137 Hz to 9037 Hz. The $\chi'(T)$ curves did not show any peak up to 340 K, although magnitude of $\chi'(T)$ decreased with the increase of frequency. The $\chi''(T)$ curves showed a peak at temperature ($T_f$) above 250 K at lower frequency (137 Hz). The peak is slowly transformed into a shoulder with increasing magnitude of $\chi''(T)$ curves at higher frequencies. The ac susceptibility data (Fig. 6(c-d)) for NFpH8_500 sample also showed dispersion on increasing the driving frequency (*f*) from 37 Hz to 9037 Hz. The $\chi'(T)$ curves did not show any peak up to 340 K, although its magnitude decreased with the increase of frequency. The $\chi''(T)$ curves showed a well defined peak at temperature ($T_f$) above 200 K with increasing magnitude on increasing the frequency of ac field. The position ($T_f$) of $\chi''(T)$ peak in both the samples shifted to higher temperature with the increase of driving frequency and the peak shift can be fitted with a general form of Vogel–Fulcher law [27, 34].

$$f(T_f) = f_o exp\left[\frac{-E_a}{k_B(T_f-T_o)}\right] \quad (7)$$

The constant $T_o$ is an effective temperature ($< T_m$) that takes into account the interaction in spin glass system and it is zero for non-interacting particles/clusters (superparamagnetic system). The $T_f(f)$ data for non-interacting superparamagnetic particles follow Néel–Arrhenius law ($f = f_o exp(-E_a/k_B T_f)$) with $f_o$ in the range $10^9$–$10^{12}$ Hz and $T_0 = 0$ K. The $\ln f$ vs. $1/(T_f-T_0)$ plot in the inset of Fig. 6(b) suggests that $T_f(f)$ data for NFpH6_500 sample are best fitted with characteristic frequency $f_0 \sim 1.8\times10^9$ Hz, $T_0 \sim 188$ K and activation energy $E_a \sim 116$ meV. The $f(T_f)$ data for NFpH8_500 sample (inset of Fig. 6(d)) are best fitted with $f_0 \sim 1.2\times10^{15}$ Hz, $T_0 = 0$ K and $E_a \sim 564$ meV. On the other hand, increase of the annealing temperature to 1000 $^0$C (NFpH6_1000) for the material prepared at pH 6 showed a significant change in ac susceptibility character. The $\chi'(T)$ curves (Fig. 7(a)) showed a prominent shoulder at about 45 K ($T_m$). The shoulder shifts to higher temperature on increasing the frequency from 37 Hz to 9037 Hz with a noticeably divergence of $\chi'(T)$ curves (decreasing magnitude) at lower temperatures. In contrast, the $\chi''(T)$ curves (Fig. 7(b)) showed a sharp peak at the inflection point ($T_f$) of the $\chi'(T)$ curves below the respective $T_m$. The divergence of $\chi''(T)$ curves decreased at higher temperature and showed a minor increasing trend with temperature, unlike a steady increase in $\chi'(T)$ curves. The frequency dependence of $T_m$, as estimated from the shoulder of $\chi'(T)$, followed equation (7) with substitution of $T_f$ by $T_m$ (Fig. 7(c)), and provided the fit parameters $f_0 \sim 2.8\times10^6$ Hz, $T_0 \sim 0$ K and $E_a \sim 40$ meV. These fit parameters are close to the values ($f_0 \sim 2.7\times10^7$ Hz, $T_0 = 0$ K and $E_a \sim 38$ meV) obtained by fitting the $T_f(f)$ data from $\chi''(T)$ peaks (Fig. 7(c)). The observed $f_0$ and $E_a$ from the low temperature shoulder of $\chi'(T)$ curves or peaks of $\chi''(T)$ curves in NFpH6_1000 sample is noticeably high in comparison to the values ($f_0 \sim 10^5$ Hz, $E_a \sim 6.5$ meV) reported for spin glass like transition at about 35 K in $Fe_3O_4$ nanoparticles [35]. The peak temperature shift per decade of frequency change ($\frac{\Delta T_f}{T_f \Delta \log(f)}$) is found $\sim$ 0.058, 0.087, and 0.293 for the samples NFpH6_500, NFpH8_500, and NFpH6_1000, respectively. It is discussed in literature [6, 23, 27, 36-37] that the typical value of $\frac{\Delta T_f}{T_f \Delta \log(f)}$ is expected to be very small ($\sim$ 0.001) for classical spin-glass, extremely large ($\sim$ 0.1) for superparamagnet, and intermediate for insulating spin-glass (EuSr)S ($\sim$0.06) and cluster spin glass ($\sim$0.032-0.044) systems. The $f_0$ values are expected in the range $10^{12}$ Hz-$10^{14}$ Hz with non-zero $T_0$ for a typical spin glass. On the other hand, $f_0$ can be expected in the range $10^6$ Hz-$10^7$ Hz for cluster spin glass systems, where spin dynamics is relatively slow due to intra-spin interactions inside the clusters [23]. The observed From application point of view, the magnetic nanoparticles with low temperature superparamagnetic blocking phenomenon and spin relaxation time at room temperature of the order of $10^{-14}$ s are essential for biomedical applications and fast computing process [38]. On the other hand, data storage medium and hyperthermia applications need magnetic nanoparticles with reasonably stable magnetization, large squareness, and controlled inter-particle interaction with blocking/ freezing temperature well above 300 K [25]. The results of the present work can be interesting for designing the ferrite nanomaterials for specific applications.

## 4. Summary and conclusions

The present work shows that ferromagnetic properties and random surface spin freezing phenomena in Ni rich ferrite ($Ni_{1.5}Fe_{1.5}O_4$) particles can effectively be engineered by adopting a combined technique of varying pH value during chemical reaction and annealing temperature. We have carried out a detailed study for the selected samples prepared at pH 6, 8 and 12. The random surface spin freezing or quantization of spin-wave spectrum in nanoparticles can be attributed as the cause of an abrupt increase of saturated magnetization below 50 K; resulted in a deviation from Bloch law of magnetic spin wave theory that obeyed for saturated magnetization of the samples at temperature $\geq$ 50 K. However, the temperature exponent ($\alpha$) is largely deviated from 3/2, a typical value assigned according to mean field theory for long ranged ferromagnet. It is understood that magnetic spin interactions are sufficiently strong (large $\alpha$ and small Bloch constant $B$) for the samples with larger grain size (prepared at pH 6 and 8) and the spin exchange interaction is highly perturbed (small $\alpha$ and large $B$) for the sample with smaller grain size (prepared at pH 12). This resulted in a decrease of ferromagnetic component and increase of superparamagnetic component in saturated magnetization along with low coercivity, low ferromagnetic squareness and low magnetic blocking temperature for the NFpH12_800 sample. The increase of surface spin disorder (superparamagnetic component) in ferromagnetic particles of the samples prepared at higher pH value is also confirmed from the observation of decreased $k$ value, obtained from fitting of $H_C(T)$ data, and ferromagnetic squarenes in our samples with the increase of pH during chemical reaction. In the present system, we have observed an interesting change in the pattern of random spin freezing behavior. Among the three samples used for ac susceptibility study, the sample (NFpH6_500) prepared at low pH value showed relatively high values of magnetization, spin freezing temperature and non-zero value of interaction parameter $T_0$ in comparison to the sample (NFpH8_500) prepared at pH 8 with interaction parameter $T_0 = 0$. The obtained values of $f_o$ suggest a relatively slow spin dynamics (interacting spin freezing phenomena) in NFpH6_500 sample in comparison to fast (non-interacting) spin freezing dynamics in NFpH8_500 sample. The increase of annealing temperature of the sample prepared at pH 6 to 1000 $^0$C (NFpH6_1000) further slowed down the spin dynamics and confined to lower temperature with reduced activation energy in comparison to the same material annealed at 500 $^0$C (NFpH6_500). The striking difference is that $T_f(f)$ data in ac susceptibility spectrum of NFpH6_500 sample are best fitted by $T_0 = 188$ K in comparison with $T_0 = 0$ K for NFpH6_1000 sample. This implies that large amount of inter-spin/particle interactions exist in the sample with low annealing temperature and thermal annealing of the sample brings a major change in the mechanism of spin freezing phenomena. In case of NFpH6_1000 sample, it is understood that small magnetic domains/grains are clustered due to increase of annealing temperature of the material. The large-sized clusters responded like non-interacting superparamagnetic particles; each of them consisting of multi-domains/grains, in the temperature and field dependence of magnetization and ac susceptibility features, and spin dynamics becomes relatively slow due to strong intra-spin interactions inside the clusters in comparison to inter-clusters interactions.

**Acknowledgment**

The authors thank CIF, Pondicherry University, for dielectric properties measurements. RNB thanks to UGC for supporting research Grant (F.No. 42-804/2013 (SR)) for the present work.

**Figure descriptions:**

Fig. 1 (a-i) Temperature dependence of MZFC(black) and MFC(red) curves at 100 Oe for the samples prepared at pH 6, 8 and 12 (lattice parameter and grain size inside the bracket). The fitting of MZFC curve for some of the samples are shown (e, h) where distribution function $f(t_p)$ was obtained by fitting the peak of the temperature derivative of MZFC curve (dMZFC/dT), which is around the inflection point of the MZFC(T) curve below blocking temperature, $T_B$.

Fig. 2 MZFC(T) curves at different magnetic fields (a-c) are fitted (lines) with a guidance to peak shift. The first order derivatives of MZFC(T) curves and distribution curves used for fitting of MZFC(T) data are shown for two samples (f, h).

Fig. 3 (a-c) Variation of the fit parameters from first order derivative of MZFC curves for two samples with magnetic field (a-c). The field dependence of blocking temperature (d). Fitted data of $T_B$ (e) and $T_{Bm}$ (f) with the expression $T_B = a - bH^n$.

Fig. 4 M(H) loops at 10 K and 300 K for the samples prepared at pH 6 and annealed at different temperatures (a-d). Inset of (b) shows the ZFC and FC loops at 10 K. The M(H) loops at different measurement temperatures are shown for 3 samples (e-g). Inset of (f-g) shows the Arrot plot. The inset of (h) shows the increment of spontaneous magnetization with annealing temperature. The fit of initial M(H) curves using law of approach to saturation of magnetization is shown for two samples (h-i).

Fig. 5 The fit of saturated magnetization of the ferromagnetic component of the samples with Bloch law (a). The inset of (a) plots the temperature dependence of the saturated magnetization due to superparamagnetic component for NFpH12_800 sample. Fit of the temperature dependence of $H_C$ with power law (b). Linear fit (y = a +bx) of the $H_C$ and $K_{eff}$ at low measurement temperature (10 K for pH 6 and 5 K for pH 8) with inverse of grain size of the samples with fit parameters in the box. The fitted curves are shown by dotted lines.

Fig. 6 Temperature dependence of ac susceptibility data for samples NFpH6_500 (a-b) and NFpH8_500 (c-d), measured at different frequencies. The frequency shift of the peak position ($T_f$) in $\chi''(T)$ data are fitted with Vogel-Fulcher law and shown as inset figures.

Fig. 7 Temperature dependence of ac susceptibility ($\chi'$ and $\chi''$) data for NFpH6_1000 sample, measured at different frequencies (a-b). The frequency shift of shoulder in $\chi'(T)$ and peak in $\chi''(T)$ data are fitted with Arrhenius law (c-d).

**Table 1**

The ferromagnetic parameters of the samples NFpH6_600, NFpH8_500 and NFpH12_800 were obtained by analyzing the MZFC(T) curves using equation (2) at different applied magnetic field (first 3 columns), and by analyzing the M(H) curves at different measurement temperatures (last 6 columns) and using equation (4).

| H(kOe) | C (emu/g) | $K_{eff}/H$ (emu/g) | Sample | T (K) | $M_S$ (emu/g) | $M_{sat}$ (emu/g) | $H_c$(Oe) | $\frac{M_R}{M_S}$ | $K_{eff}$ (emu-Oe/g) |
|---|---|---|---|---|---|---|---|---|---|
| 0.10 | 0.65 | 527 | NFpH6_600 | 10 | 28.6 | 34.1 | 532.5 | 0.46 | 542019 |
| 0.25 | 1.02 | 812 | | 50 | 27.1 | 33.4 | 505 | 0.47 | 441587 |
| 0.50 | 1.53 | 788 | | | | | | | |
| 1.00 | 3.75 | 917 | | 100 | 26.76 | 32.8 | 360 | 0.33 | 405095 |
| 1.50 | 7.90 | 1195 | | 150 | 26.4 | 32.9 | 283.5 | 0.23 | 482310 |
| 2.50 | 11.20 | 3920 | | 200 | 25.4 | 31.8 | 141.5 | 0.15 | 427374 |
| 5.00 | 12.90 | 4968 | | 250 | 24.6 | 30.7 | 65.6 | 0.09 | 435604 |
| 10.00 | 15.14 | 26790 | | 300 | 23.9 | 28.9 | 21.16 | 0.02 | 334114 |
| 0.10 | -0.34 | 1615 | NFpH8_500 | 5 | 22.6 | 30.1 | 884 | 0.44 | 582338 |
| 0.20 | 0.56 | 1186 | | 50 | 22.38 | 28.6 | 631 | 0.38 | 503527 |
| 0.50 | 1.23 | 635 | | 100 | 22.24 | 28.6 | 357 | 0.21 | 540937 |
| 0.75 | 2.37 | 545 | | 150 | 21.9 | 27.9 | 168 | 0.11 | 498838 |
| 1.00 | 2.54 | 435 | | 200 | 20.8 | 27.8 | 119 | 0.15 | 549206 |
| 1.50 | 4.58 | 419 | | 250 | 20.7 | 26.4 | 74.9 | 0.06 | 454682 |
| 2.00 | 5.97 | 399 | | 300 | 19.7 | 25.1 | 7.08 | 0.01 | 23897 |
| 3.00 | 9.39 | 570 | | | | | | | |
| 0.10 | 0.19 | 2246 | NFpH12_800 | 5 | 13.2 | 16.5 | 1122 | 0.28 | 239704 |
| 0.20 | 0.45 | 1184 | | 25 | 9.73 | 14.8 | 514 | 0.25 | 160406 |
| 0.50 | 0.92 | 534 | | 50 | 9.59 | 12.8 | 160 | 0.11 | 101490 |
| 0.75 | 1.33 | 396 | | 75 | 9.37 | 12.4 | 27.5 | 0.028 | 101768 |
| 1.00 | 1.76 | 332 | | 100 | 7.74 | 11.9 | 6.66 | 0.008 | 101142 |
| 1.50 | 3.66 | 355 | | 150 | 6.4 | 10.6 | 19.67 | 0.018 | 95079 |
| 3.00 | 5.52 | 286 | | 200 | 4.29 | 9.28 | - | - | 89789 |
| | | | | 250 | 1.86 | 7.86 | | | 80804 |
| | | | | 300 | | 6.15 | | | 62702 |

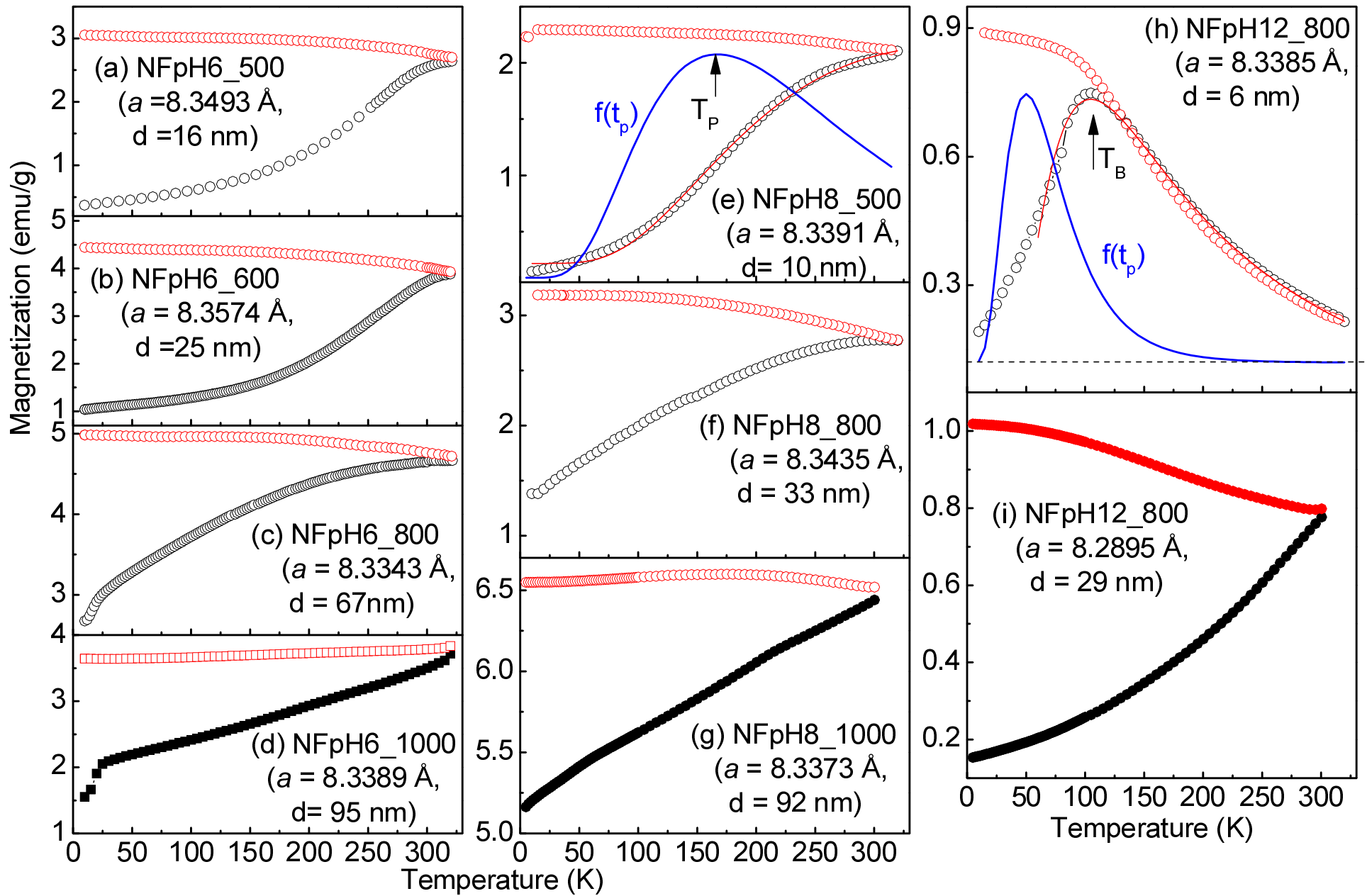

Fig. 1(a-i) Temperature dependence of MZFC(black) and MFC(red) curves at 100 Oe for the samples prepared at pH 6, 8 and 12 (lattice parameter and grain size inside the bracket). The fit of MZFC curve for some of the samples are shown (e, h) where distribution function $f(t_p)$ was obtained by fitting the peak of temperature derivative of MZFC curve (dMZFC/dT), which is around the inflection point of the MZFC(T) curve below blocking temperature, $T_B$.

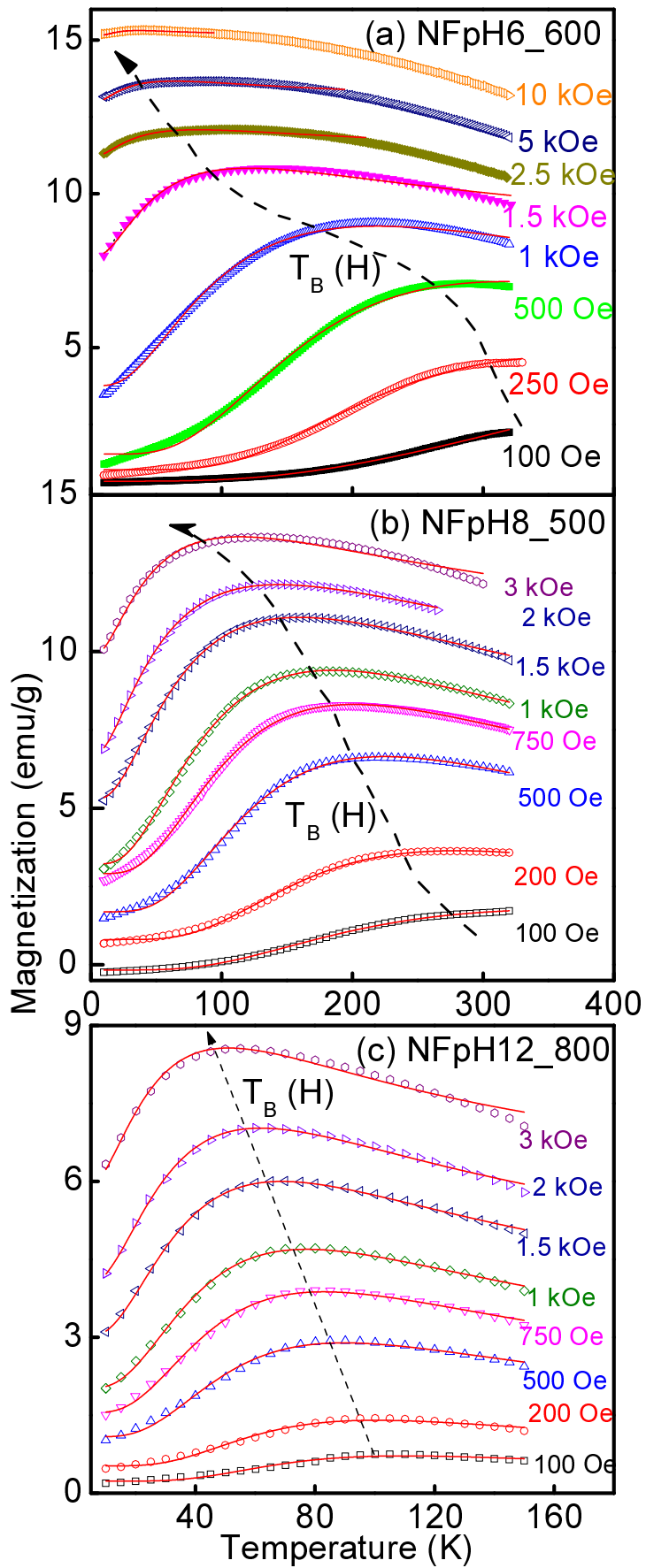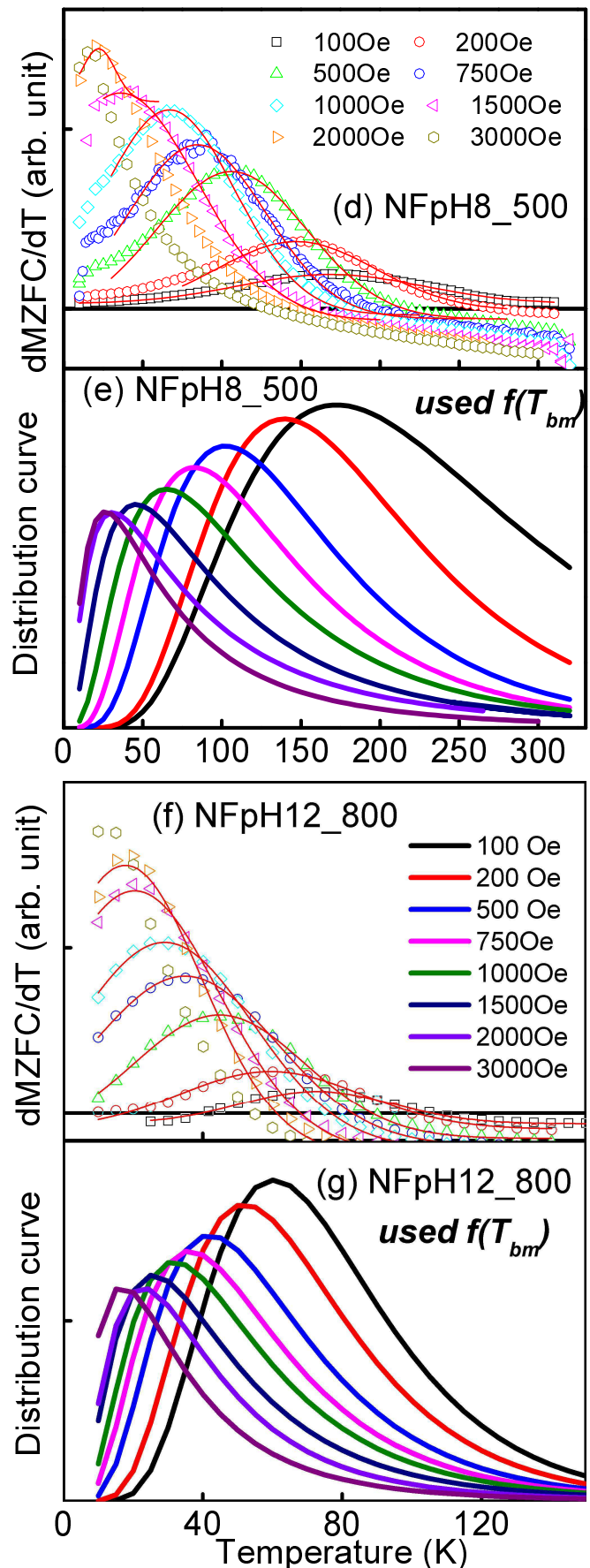

Fig. 2 MZFC(T) curves at different magnetic fields (a-c) are fitted (lines) with a guidance to peak shift. The first order derivatives of MZFC(T) curves and distribution curves used for fitting of MZFC(T) data are shown for two samples (f, h).

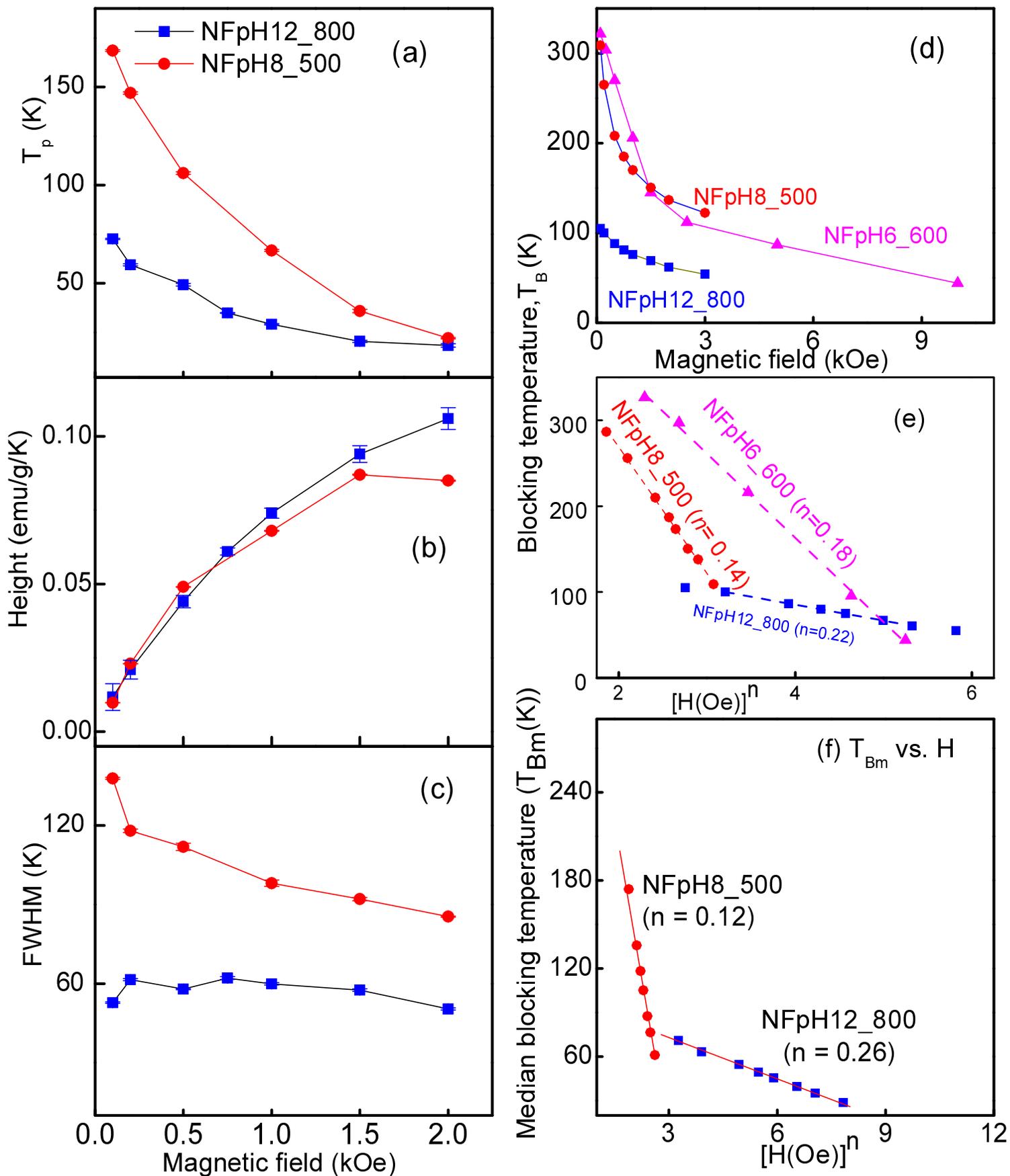

Fig. 3 (a-c) Variation of the fitt parameters from first order derivative of MZFC curves for two samples with magnetic field (a-c). The field dependence of blocking temperature (d). Fitted data of $T_B$ (e) and $T_{Bm}$ (f) with the expression $T_B = a - bH^n$.

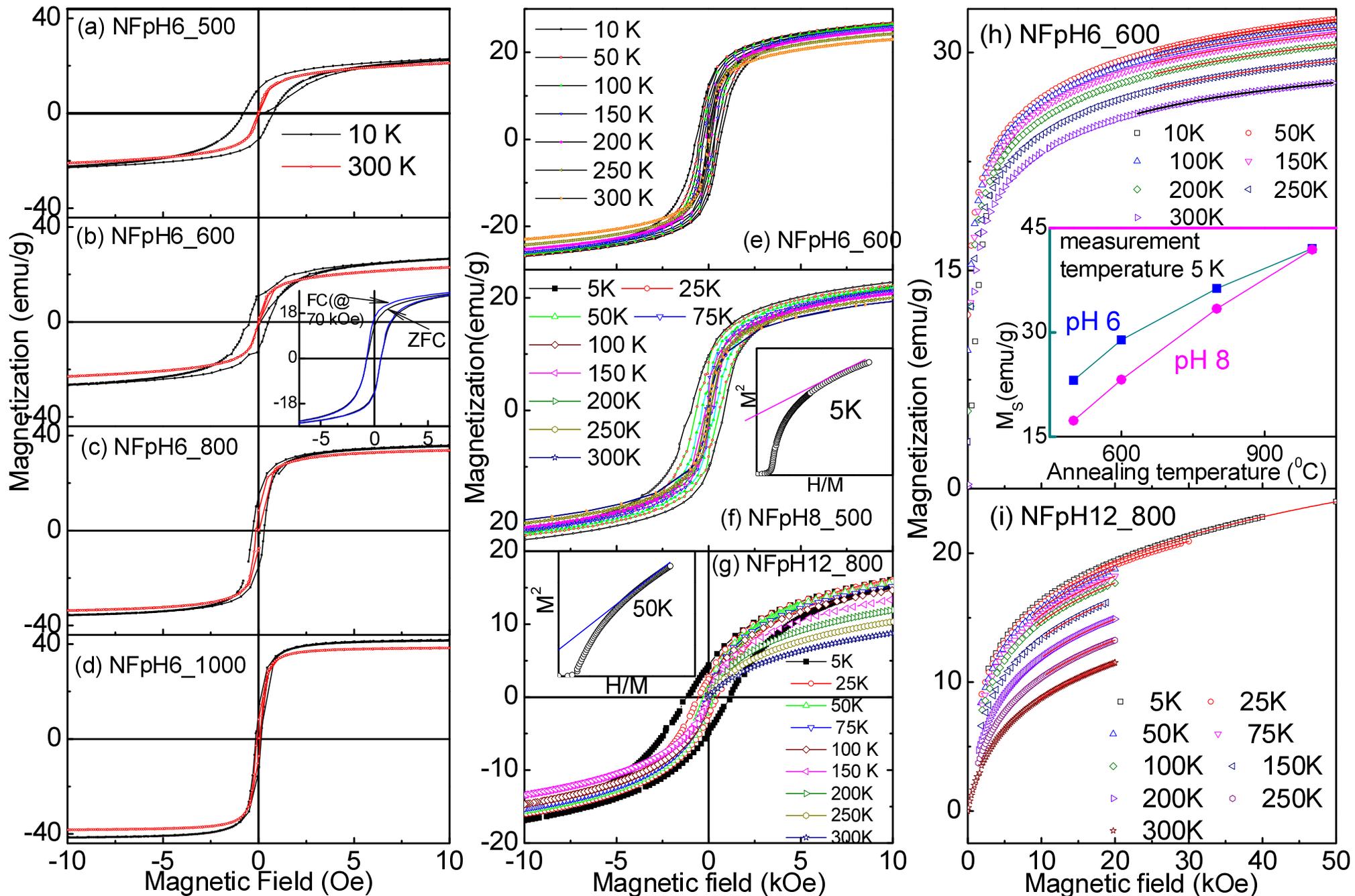

Fig. 4 M(H) loops at 10 K and 300 K for the samples prepared at pH 6 and annealed at different temperatures (a-d). Inset of (b) shows the ZFC and FC loops at 10 K. The M(H) loops at different measurement temperatures are shown for 3 samples (e-g). Inset of (f-g) shows the Arrot plot. The inset of (h) shows the increment of spontaneous magnetization with annealing temperature. The fit of initial M(H) curves using law of approach to saturation of magnetization is shown for two samples (h-i).

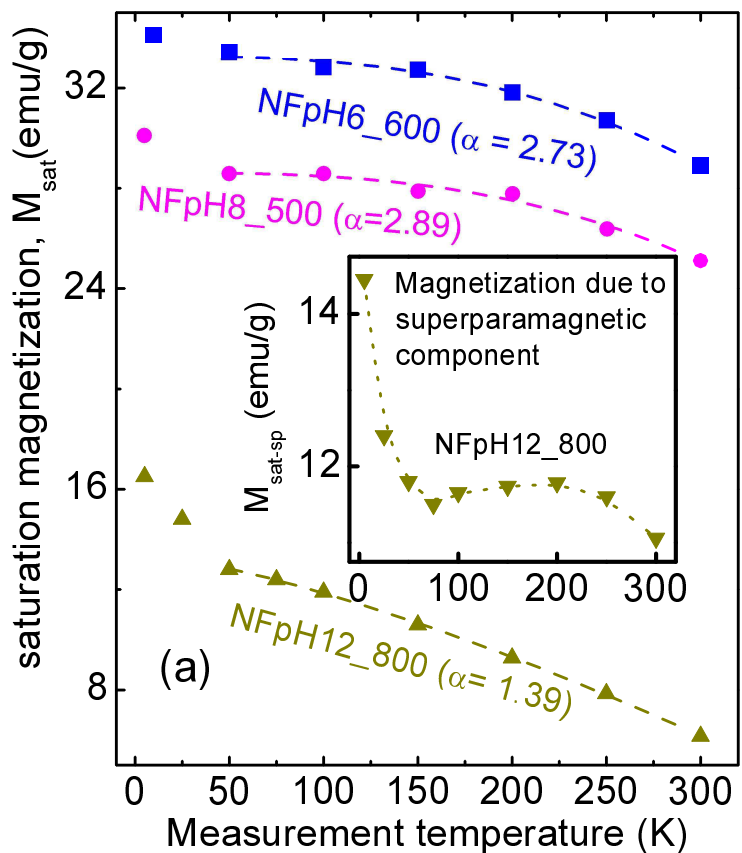
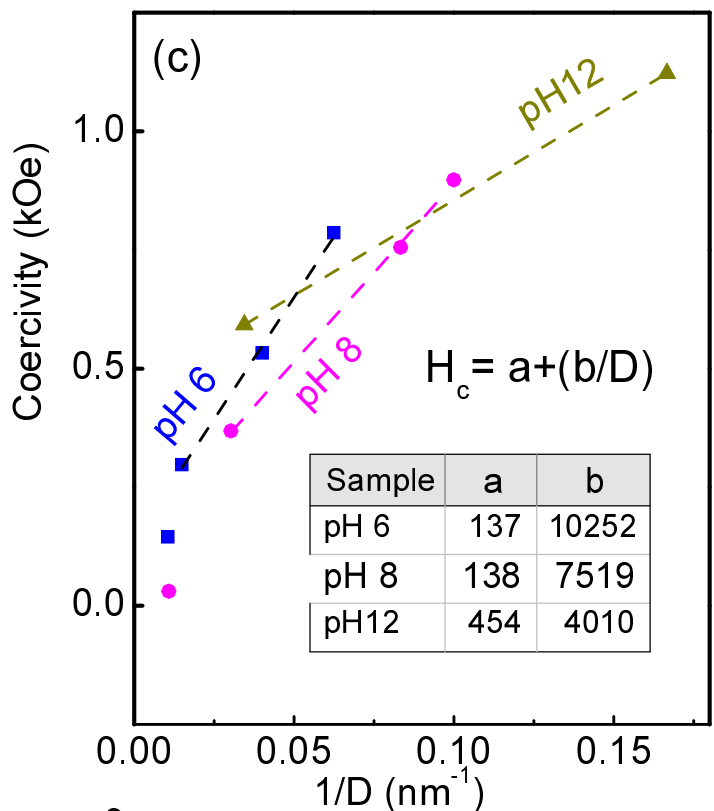
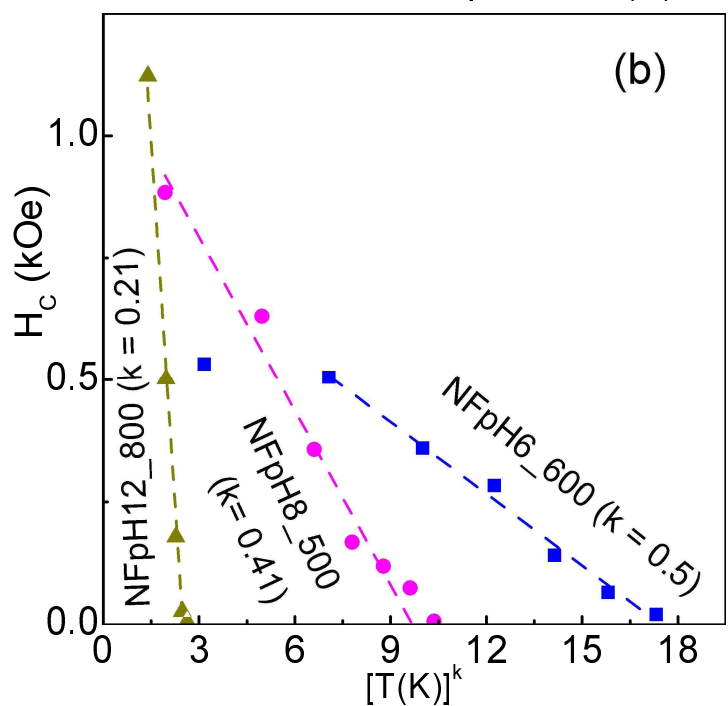
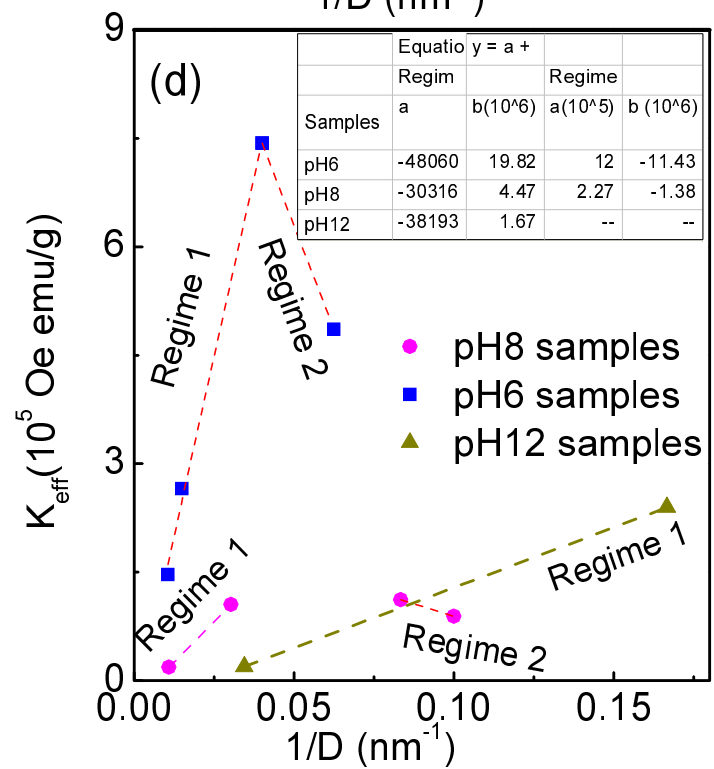

Fig. 5 The fit of saturated magnetization of the ferromagnetic component of the samples with Bloch law (a). The inset of (a) plots the temperature dependence of the saturated magnetization due to superparamagnetic component for NFpH12_800 sample. Fit of the temperature dependence of $H_c$ with power law (b). Linear fit ($y = a + bx$) of the $H_c$ and $K_{eff}$ at low measurement temperature (10 K for pH 6 and 5 K for pH 8) with inverse of grain size of the samples with fit parameters in the box. The fitted curves are shown by dotted lines.

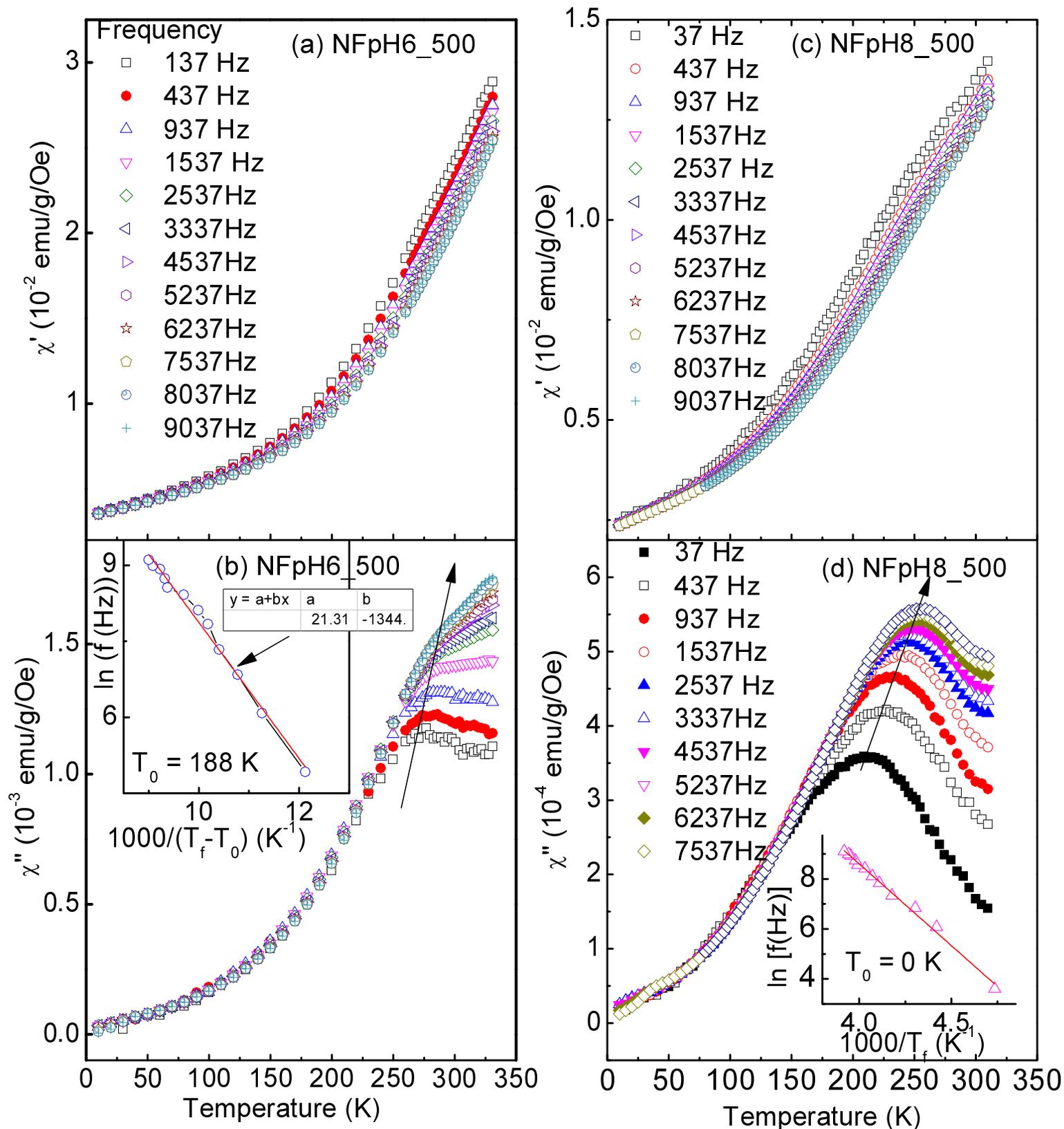

Fig. 6 Temperature dependence of ac susceptibility data for samples NFpH6_500 (a-b) and NFpH8_500 (c-d), measured at different frequencies. The frequency shift of the peak position ($T_f$) in $\chi''(T)$ data are fitted with Vogel-Fulcher law and shown as inset figures.

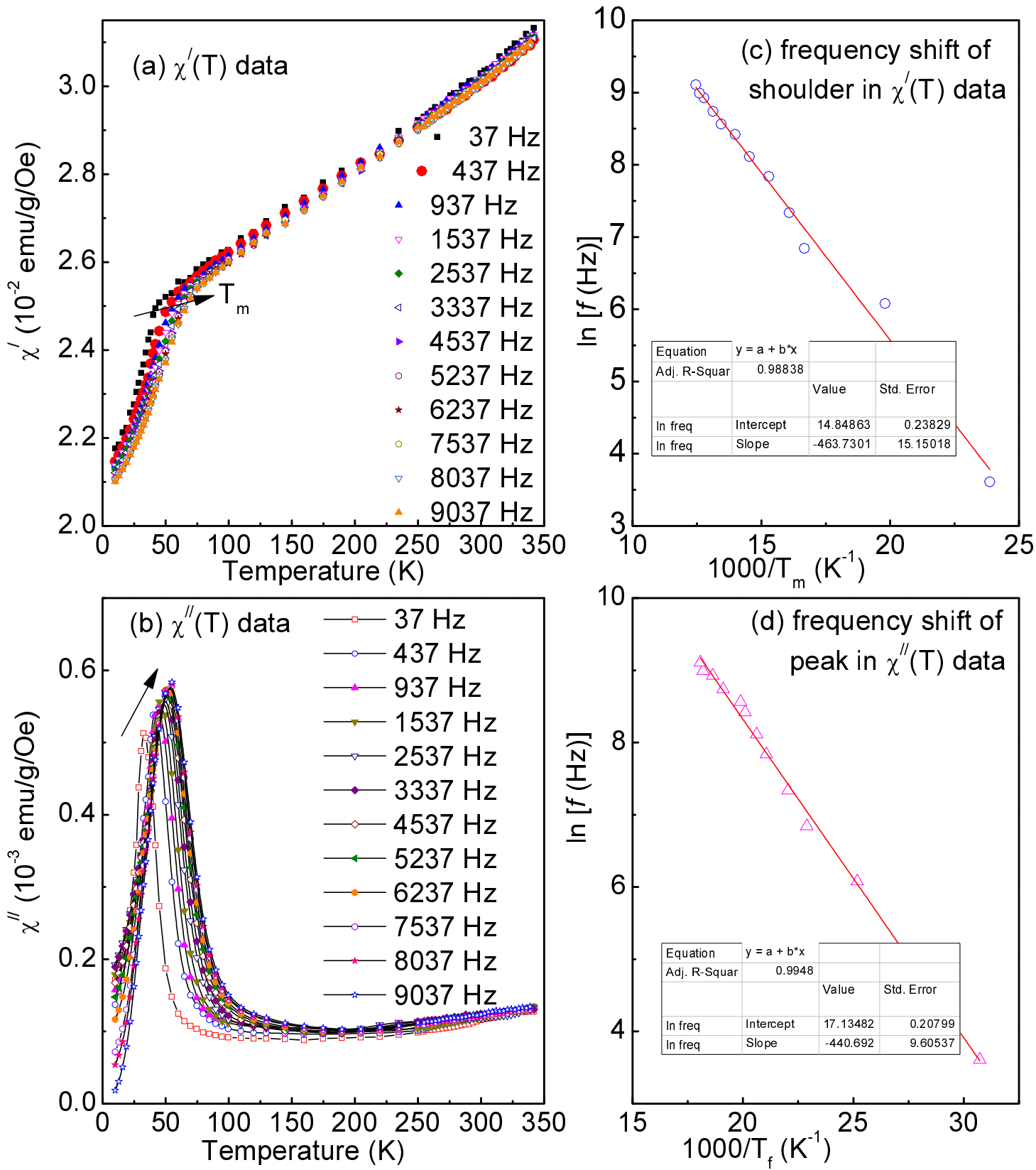

Fig. 7 Temperature dependence of ac susceptibility ($\chi'$ and $\chi''$) data for NFpH6_1000 sample, measured at different frequencies (a-b). The frequency shift of shoulder in $\chi'(T)$ and peak in $\chi''(T)$ data are fitted with Arrhenius law (c-d).